\documentstyle{article}

\input epsf.sty

\newcommand{\beq}{\begin{equation}}
\newcommand{\eeq}{\end{equation}}
\newcommand{\ba}{\begin{array}}
\newcommand{\ea}{\end{array}}
\newcommand{\bea}{\begin{eqnarray}}
\newcommand{\eea}{\end{eqnarray}}
\newcommand{\bean}{\begin{eqnarray*}}
\newcommand{\eean}{\end{eqnarray*}}

\begin{document}

\begin{center}

{\Large \sc \bf INITIAL-BOUNDARY VALUE PROBLEMS}

\vskip 5pt

{\Large \sc \bf FOR LINEAR PDEs:} 

\vskip 5pt

{\Large \sc \bf THE ANALYTICITY APPROACH} 

\vskip 20pt

{\large  A. Degasperis$^{1,2,\S}$, S. V. Manakov$^{3,\S}$ and 
P. M. Santini$^{1,2,\S}$}

\bigskip

{\it
 $^1$Dipartimento di Fisica, Universit\`a di Roma "La Sapienza"\\
Piazz.le Aldo Moro 2, I-00185 Roma, Italy

\smallskip

$^2$Istituto Nazionale di Fisica Nucleare, Sezione di Roma\\
P.le Aldo Moro 2, I-00185 Roma, Italy

\smallskip

$^3$Landau Institute for Theoretical Physics, Moscow, Russia}

\bigskip

$^{\S}$e-mail: ~{\tt antonio.degasperis@roma1.infn.it,~ manakov@itp.ac.ru, \\ 
paolo.santini@roma1.infn.it}

\bigskip

\end{center}

\begin{abstract}

\noindent
It is well-known that the main difficulties associated with the study of 
initial-boundary value problems for linear PDEs  
is given by the presence of unknown boundary values in 
any method of solution. To deal efficiently with this difficulty, we have 
recently proposed two alternative (but interrelated) methods in Fourier space: 
the Analitycity approach and the Elimination by Restriction approach. In this 
work we present the Analyticity approach and we illustrate its power in 
studying the well-posedness of initial - boundary value problems for second 
and third order evolutionary PDEs, and in constructing their solution. We also 
show the connection between the Analyticity 
approach and the Elimination by Restriction approach in the study of the 
Dirichelet and Neumann problems for the Schr\"odinger equation in the 
$n$-dimensional quadrant. 
\end{abstract}

\vskip 20pt

\section{Introduction}

It is well-known that the main difficulties associated with Initial-Boundary 
Value (IBV) problems for linear PDEs of the type

\beq\label{eq:IBV}
{\cal L}({\bf \bigtriangledown},{\partial\over\partial t})u({\bf x},t)=
f({\bf x},t),~~u({\bf x},0)=u_0({\bf x}),~~{\bf x}\in V\subset {\cal R}^n,
~~t>0,
\eeq
where ${\bf \bigtriangledown} =({\partial\over\partial {x_1}},\cdot\cdot ,
{\partial\over\partial {x_n}})$, ${\cal L}$ is a constant coefficients 
partial differential operator, $u({\bf x},t)$ is the unknown field, 
$f({\bf x},t)$ is a given forcing and $u_0({\bf x})$ is the given initial 
condition, with Dirichelet, or Neumann, or Robin, or mixed, or periodic 
boundary data on $\partial V$, 
is given by the presence of unknown Boundary Values (BVs) in 
any method of solution. To deal efficiently with this difficulty, we have 
recently proposed two alternative (but interrelated) methods in Fourier space: 
the {\it Analyticity approach} and the {\it Elimination by Restriction 
approach}. 
 
The first step, common to both methods,  consists in rewriting the PDE 
(\ref{eq:IBV}), 
defined in a space-time domain $\cal D$, in the corresponding Fourier space, 
using the Green's formula. The PDE in Fourier space takes the form of a   
linear relation among the Fourier Transforms (FTs) of the solution,  
of the initial condition and of a set of BVs, only a subset of which 
is given a priori. This relation is always supplemented by 
strong analyticity requirements on all the FTs involved, consequence of the 
geometric properties of the space-time domain $\cal D$.   

The second step is where the two methods separate; 
once the problem is formulated in Fourier space, we propose the 
following two alternative strategies.

\noindent
i) The {\it Analyticity approach}, which consists in using systematically 
the analyticity properties 
of all the FTs involved in the above relation, to derive a 
system of linear equations which allows 
one to express the unknown BVs in terms of the known ones, and 
therefore to solve the problem.

\noindent
ii) {\it The Elimination by Restriction (EbR) approach}, which consists, 
instead, in applying to the above linear relation in Fourier space a 
suitable annihilation operator, 
which eliminates all the unknown BVs, generating a new transform, 
well-suited to the specific IBV problem under scrutiny. The inversion of this 
new transform (if it exists) leads to the solution. 

The Analyticity approach is inspired by Fokas' recent discovery of 
the global relation, obtained first 
within the $x-t$ transform approach \cite{Fokas1} and more recently using 
differential forms \cite{Fokas2}. The use of the global relation to study 
the well-posedness and solve 
IBV problems is illustrated, for instance, in \cite{Fokas3}, \cite{FP},  
\cite{Pelloni}. In \cite{Pelloni}, in particular, general results on the 
well-posedness of IBV problems for dispersive $1+1$ dimensional equations 
of arbitrary order are announced.

Our main contribution to the method consists,  
after formulating the IBV problem in Fourier space using Green's formula, 
in imposing systematically the analyticity properties of all the Fourier 
transforms involved in the problem, 
 to derive a cascade of analyticity constraints which 
allow one to express the unknown BVs in terms of the known ones, and 
therefore to solve the problem. In particular, Fokas' global relation appears, 
in the methodology we propose, as a ``zero residue condition'' for the 
FT of the solution.
 
The Analyticity approach in the form we propose is very elementary and, above all, 
has the great conceptual advantage to originate from a single 
guiding principle: the analyticity of all the Fourier transforms 
involved in the problem. It is the type of approach that can be easily 
taught in elementary University courses, combining nicely standard 
PDE theory tools, like the Green's formula and the Fourier transform, with 
elementary notions in Complex Functions theory. 

The essential aspects of the Analyticity approach were first presented by 
the authors at the Workshop ``Boundary value problems'' in Cambridge, 
December 2001, inside the programme: ``Integrable Systems''. The method in its 
final form is presented for the first time in this paper,   
illustrated on the study of IBV problems of various type (Dirichelet, Neumann, 
mixed, periodic) for 
some second and third order classical PDEs of 
the Mathematical Physics: the Schr\"odinger, the heat and the linear Korteweg-de 
Vries equations.  
Also its connections with the EbR approach are 
illustrated in this work, on the particular example of the Schr\"odinger 
equation in the $n$-dimensional quadrant. A good account of the EbR approach 
is given instead in \cite{DMS1}. A different approach, valid for semicompact 
domains, has been recently presented in \cite{Fokas4}. A general 
review of the basic spectral methods of solution of IBV problems for linear 
and soliton PDEs is presented in \cite{DMS2}. 

\S 2 is devoted to the presentation of the Analyticity approach, 
while \S 3 is dedicated to its  
application to the solution of some IBV problems for second and third order 
evolutionary PDEs in $1+1$ and in $n+1$ dimensions. \S 4 is finally devoted 
to the study of the connections between the Analyticity approach and the EbR 
approach.

\section{The Analyticity Approach}

\subsection{The Fourier transform and its analiticity properties}

The natural FT associated with the space - time domain 
${\cal D}=V\otimes (0,\infty )$ (in short: $FT_{\cal D}$) is defined by
\beq\label{eq:FT}
\hat F({\bf k},q)=\int_{\cal D}d{\bf x}dt
e^{-i({\bf k}\cdot {\bf x}+qt)}F({\bf x},t)
\eeq 
for any smooth function $F({\bf x},t)$, $({\bf x},t)\in {\cal D}$, assuming 
that $F({\bf x},t)\to 0,~t\to\infty$ fast enough; where 
${\bf k}=(k_1,..,k_n)\in {\cal R}^{n},~q\in {\cal R}$ and 
${\bf k}\cdot {\bf x}=\sum_{j=1}^nk_jx_j$. Its inverse:
\beq\label{eq:invFT} 
F({\bf x},t)\chi_{\cal D}({\bf x},t)=
\int_{{\cal R}^{n+1}}{d{\bf k}dq\over (2\pi )^{n+1}}
e^{i({\bf k}\cdot {\bf x}+qt)}\hat F({\bf k},q)
\eeq
reconstructs $F({\bf x},t)$ in $\cal D$ and zero outside, 
where $\chi_{\cal D}({\bf x},t)$ is the characteristic 
function of the domain $\cal D$: $\chi_{\cal D}({\bf x},t)=1,~({\bf x},t)\in 
{\cal D},~\chi_{\cal D}({\bf x},t)=0,~({\bf x},t)\notin {\cal D}$ 
(therefore: $\chi_{\cal D}({\bf x},t)=
\chi_{V}({\bf x})H(t)$, where $H(t)$ is the usual Heaviside (step) 
function). 

If the space domain is the whole space: $V={\cal R}^n$, the $FT_{\cal D}$ 
(\ref{eq:FT}) is defined in ${\cal A}={\cal R}^n\otimes \bar{\cal I}_q$, where 
$\bar{\cal I}_q$ is the closure of the lower half $q$-plane ${\cal I}_q$, 
analytic in $q\in {\cal I}_q,~\forall 
{\bf k}\in {\cal R}^n$ and exhibits a proper asymptotic 
behaviour for large $q$ in the analyticity region.  
If the space domain $V$ is compact, the $FT_{\cal D}$ acquires strong 
analyticity properties in all the Fourier variables: it is defined in 
${\cal A}={\cal C}^n\otimes\bar{\cal I}_q$,
 analytic in $q\in {\cal I}_q,~\forall {\bf k}\in {\cal C}^n$, entire  
in every complex $k_j,~j=1,..,n$ $\forall q\in \bar{\cal I}_q$ and  
exhibits a proper asymptotic behaviour, for 
large $({\bf k},q)$, in the analyticity regions.  
If the space domain is semi - compact, then the analyticity in the Fourier 
variables $k_j,~j=1,..,n$ is limited to open regions of the complex plane, 
depending on the geometric properties of the domain $V$.

\noindent
We are therefore led to the following definition:

\vskip 5pt
\noindent
{\bf Definition of admissibility}.  
Given a space-time domain $\cal D$, a function of $({\bf k},q)$ is an 
{\bf admissible Fourier transform for the domain $\cal D$} (an 
admissible $FT_{\cal D}$) iff it possesses 
the analyticity properties and the asymptotic behaviour corresponding to that 
domain.

\subsection{The IBV problem in Fourier space}

We find it convenient to rewrite the IBV problem (\ref{eq:IBV}) in Fourier 
space. 
This goal is conveniently achieved using the well - known {\bf Green's formula 
(identity)}:

\beq\label{eq:Gformula}
b{\cal L}a-a \tilde {\cal L} b=div ~J({\bf x},t),
\eeq
and its integral consequence, the celebrated {\bf Green's integral 
identity}:
\beq\label{eq:Gint}
\int_{{\cal D}}(b{\cal L}a-a \tilde {\cal L} b)d{\bf x}dt=
\int_{\partial {\cal D}}J({\bf x},t)\cdot {\bf \nu} d\sigma , 
\eeq
obtained by integrating (\ref{eq:Gformula}) over the domain ${\cal D}$ and 
by using the divergence theorem. In equation (\ref{eq:Gformula}),  
$\tilde {\cal L}$ is the formal adjoint of ${\cal L}$:  
$\tilde {\cal L}= {\cal L}(-{\bf \bigtriangledown},
-{\partial\over\partial t})$, $J({\bf x},t)$ is an $(n+1)$-dimensional 
vector field, 
$div$ is the $(n+1)$-dimensional divergence operator and     
$a({\bf x},t)$ and $b({\bf x},t)$ are arbitrary functions.  In equation 
(\ref{eq:Gint}),   
$d\sigma$ is the hypersurface element of the boundary and $\bf \nu$ is 
its outward unit normal. We remark that, given ${\cal L}$, its 
formal adjoint $\tilde {\cal L}$ and two arbitrary functions $a$ and $b$, 
an $(n+1)$-dimensional vector field $J({\bf x},t)$ satisfying 
the Green's formula (\ref{eq:Gformula}) always exists and can be  
algorithmically found to be a linear expression of $a$, $b$ and their 
partial derivatives of order up to $N-1$, if $\cal L$ is of order $N$. 

The arbitrariness of $a$ and $b$ allows one to extract from 
(\ref{eq:Gformula}) and (\ref{eq:Gint}) several important informations on the 
IBV problem; with the particular choice  
\beq\label{eq:choice1}
a=u({\bf x},t),~~~~~~~~~b=e^{-i({\bf k}\cdot{\bf x}+qt)}/{\cal L}(i{\bf k},iq),
\eeq
where ${\cal L}(i{\bf k},iq)$ is the eigenvalue of the operator $\cal L$, 
corresponding to the eigenfunction $e^{i({\bf k}\cdot {\bf x}+qt)}$,   
the vector field $J$ takes the following form: \break
$J=e^{-i({\bf k}\cdot{\bf x}+qt)}J'({\bf x},t;
{\bf k},q)/{\cal L}(i{\bf k},iq)$ and 
the Green's integral identity (\ref{eq:Gint}) gives the $FT_{\cal D}$ 
of the solution in terms of the $FT_{\cal D}$'s (or, maybe, of 
generalized FT's) of the forcing 
and of all the IBVs:
\beq\label{eq:FTu1}
\hat u({\bf k},q)={\hat f({\bf k},q)-
\int_{\partial {\cal D}}e^{-i({\bf k}\cdot {\bf x}+qt)}
J'({\bf x},t;{\bf k},q)\cdot \nu d\sigma \over 
{\cal L}(i{\bf k},iq)}=:
{\hat{\cal N}({\bf k},q)\over {\cal L}(i{\bf k},iq)},~~~
({\bf k},q)\in{\cal A}.  
\eeq

If the PDE has the following evolutionary form:
\beq\label{eq:Levol}
{\cal L}({\bf \bigtriangledown},{\partial\over\partial t})=
{\partial\over\partial t}-{\cal K}({\bf \bigtriangledown})
\eeq
then
\beq\label{eq:FTu2}
\hat u({\bf k},q)={\hat f({\bf k},q)+\hat u_0({\bf k})+
\hat {\cal B}({\bf k},q) \over 
{\cal L}(i{\bf k},iq)}=:
{\hat{\cal N}({\bf k},q)\over {\cal L}(i{\bf k},iq)},~~~
({\bf k},q)\in{\cal A}  
\eeq
and the linear relation (\ref{eq:FTu2}) makes clear how the different 
contributions coming from the equation (the denominator $\cal L$), from the 
forcing $\hat f$, from the initial condition $\hat u_0$ and from the set of 
boundary values $\hat {\cal B}$ separate in Fourier space.
 
Its inverse transform (\ref{eq:invFT}) gives the corresponding {\bf Fourier 
representation} of the solution:
\beq\label{eq:Frepr}
U({\bf x},t)=
u({\bf x},t)\chi_{\cal D}({\bf x},t)=\int_{{\cal R}^{n+1}}{d{\bf k}dq\over 
(2\pi )^{n+1}}e^{i({\bf k}\cdot {\bf x}+qt)}
{\hat{\cal N}({\bf k},q)\over {\cal L}(i{\bf k},iq)},~~
({\bf x},t)\in{\cal R}^{n+1}. 
\eeq

Two sources of problems arise at a first glance of equation (\ref{eq:FTu1}):

\noindent
i) the RHS of the equation depends on known and unknown BVs;

\noindent
ii) apparently the RHS of the equation  
is not an admissible $FT_{\cal D}$. 

It is very satisfactory that the analyticity 
constraints which make the RHS of (\ref{eq:FTu1}) an admissible $FT_{\cal D}$ 
provide also a number of relations among the IBVs which are sufficient to 
express the unknown BVs in terms of known boundary data.

\subsection{The analiticity constraints and their resolution} 

\vskip 5pt
\noindent
In general, ${\cal L}(i{\bf k},iq)$,  
the denominator of equation (\ref{eq:FTu1}), is 
an entire and, most frequently, polynomial 
function of all its complex variables. Let $\cal S$ be the manifold in which 
this entire function is zero:
\beq\label{eq:defS}
{\cal S}=\{({\bf k},q)\in {\cal C}^{n+1}:~{\cal L}(i{\bf k},iq)=0\}. 
\eeq
Then the RHS of equation (\ref{eq:FTu1}) provides an admissible 
$FT_{\cal D}$ of the solution of the 
IBV problem under investigation if the numerator $\hat{\cal N}({\bf k},q)$ of 
$\hat u$ in (\ref{eq:FTu1}) satisfies in ${\cal A}\cap {\cal S}$, 
hereafter called the {\bf singularity manifold (SM) of the IBV problem}, the  
following {\bf Zero Residue Condition} (ZRC):
\beq\label{eq:ZRC}  
\hat{\cal N}({\bf k},q)=0,~~~~~({\bf k},q)\in{\cal A}\cap {\cal S}.
\eeq

If the singularity manifold ${\cal A}\cap {\cal S}$ contains the real axis 
(which is usually a part of the boundary of $\cal A$) and if this  
singularity is not already taken care of by the ZRC (\ref{eq:ZRC}), then 
we must also proceed to the {\bf Denominator Regularization} (DR):
\beq\label{eq:reg}
{\cal L}(i{\bf k},iq)\to {\cal L}_{reg}(i{\bf k},iq),
\eeq 
which consists in moving a bit the singularity off the real axis, outside 
the domain $\cal A$.

The ZRC plus the DR constitute the main set of {\bf Analyticity Constraints} 
(ACs) that must be imposed to the RHS of (\ref{eq:FTu1}) in order to obtain 
an admissible $FT_{\cal D}$ of the solution of the IBV problem under 
investigation. 
The ZRC (\ref{eq:ZRC}) provides a (linear) relation 
among the FTs of the forcing, of the initial condition and of all the BVs.  
The analyticity 
properties of all these FTs  generate, through the 
admissibility argument, a cascade of further analyticity constraints, untill 
all these conditions are finally met. This procedure defines, 
in principle,  a set 
of relations (a system of equations) among the IBVs. 
Therefore:

\noindent
{\bf a) The unique 
solvability of such a system,
together with the admissibility of the obtained solution, 
are equivalent to the study of the unique solvability of all the 
IBV problems associated with (\ref{eq:IBV})}.

\noindent
{\bf b) By solving this system for a set of BVs 
in terms of 
the remaining ones, one expresses all quantities in terms of known data and, 
from equation (\ref{eq:Frepr}), one obtains the Fourier representation of the 
solution}.

In most of the examples considered in this paper, this system of equations 
is algebraic, with entire coefficients. Therefore, if $M$ is 
the squared matrix of the coefficients of the unknown BVs, the admissibility 
argument imposes that the countable number of zeroes of $det~M$:
\beq\label{eq:spectrum}
\{q_m\}_{m\in{\cal N}},~~~~~~det~M(q_m)=0
\eeq
lie outside the analyticity domain of an admissible $FT_{\cal D}$:
\beq
q_m\notin {\cal I}_q,~~~~m\in{\cal N}.
\eeq
It turns out that the set (\ref{eq:spectrum}) coincides with the spectrum 
arising in the eigenfunction expansion approach \cite{MF} 
and coincides also with the restricted domain in which the EbR 
method works. These deep connections justify for (\ref{eq:spectrum}) the 
name of {\bf spectrum of the IBV problem}.
 
The admissibility argument imposes also that the constructed solution 
of the system exhibit the proper asymptotic behaviour in the 
analyticity domain. It is actually convenient to impose first this 
asymptotic admissibility, the easiest to be checked, which enables 
one to disregard without effort all the IBV problems ill-posed 
because inconpatible with asymptotics.

\subsection{General remarks}

\vskip 5pt
\noindent
{\bf Remark 1. Analyticity vs Causality}.  
It is well - known that there are definite connections 
between the analyticity properties of the FT of the 
solution of evolution equations and the causality principle. In our general 
setting it is straightforward to show that:

\noindent
{\bf The analyticity properties of the $FT_{\cal D}$ of the solution of the 
IBV problem (\ref{eq:IBV}) imply the causality principle}.

\noindent
Indeed, using the  
convolution theorem, the inverse FT (\ref{eq:Frepr}) of the RHS of 
equation 
(\ref{eq:FTu1}) (in which all the analyticity constraints have been 
preliminary imposed) is equivalent to the following Green's representation 
of the solution:
\beq\label{eq:causality}
u({\bf x},t)=\int\limits_0^tdt'\int\limits_{V}d{\bf x'}
G_{RF}({\bf x}-{\bf x}';t-t'){\cal N}({\bf x'},t'),~~({\bf x},t)\in {\cal D},
\eeq
where ${\cal N}({\bf x},t)\chi_{\cal D}({\bf x},t)$ is the inverse FT 
(\ref{eq:invFT})  
of $\hat{\cal N}({\bf k},q)$ and $G_{RF}$ is the celebrated retarded - 
fundamental Green's function of the operator $\cal L$:
\beq\label{eq:GRF}
G_{RF}({\bf x},t)=\int_{{\cal R}^{n+1}}{d{\bf k}dq\over (2\pi )^n}
{e^{i({\bf k}\cdot{\bf x}+qt)}\over 
{\cal L}_{reg}(i{\bf k},iq)},
\eeq
which satisfies the important property: $G_{RF}({\bf x},t)=0,~t<0$, 
due to the regularization of 
${\cal L}(i{\bf k},iq)$. Equation (\ref{eq:causality}) is the usual way in 
which the causality principle becomes transparent.

\vskip 5pt
\noindent
{\bf Remark 2. Regularization and Fourier representation}. As we have already 
written, if the zeroes of the denominator on the real axis  
are all cured by the ZRC, no regularization is needed. On the other hand, 
some regularization must be introduced also in this case, in the 
calculation the Fourier representation (\ref{eq:Frepr}), before splitting  
$\hat{\cal N}$ in the sum of the different contributions (each one singular on 
the real axis) in (\ref{eq:FTu2}) coming from the forcing, from 
the initial condition and from the BVs. The most convenient 
regularization is obviously that in (\ref{eq:reg}) and leads to the following 
Fourier representation:
\bea\label{eq:Frepr2}
\ba{c}
U({\bf x},t)=u({\bf x},t)\chi_{\cal D}({\bf x},t)=
\int_{{\cal R}^{n+1}}{d{\bf k}dqe^{i({\bf k}\cdot {\bf x}+qt)}\over 
(2\pi )^{n+1}}
{\hat f({\bf k},q)\over {\cal L}_{reg}(i{\bf k},iq)}+ \\
\int_{{\cal R}^{n+1}}{d{\bf k}dqe^{i({\bf k}\cdot {\bf x}+qt)}\over 
(2\pi )^{n+1}}
{\hat u_0({\bf k})\over {\cal L}_{reg}(i{\bf k},iq)}+
\int_{{\cal R}^{n+1}}{d{\bf k}dqe^{i({\bf k}\cdot {\bf x}+qt)}\over 
(2\pi )^{n+1}}
{\hat {\cal B}({\bf k},q)\over {\cal L}_{reg}(i{\bf k},iq)},~~~
({\bf x},t)\in{\cal R}^{n+1}. 
\ea
\eea

\vskip 5pt
\noindent

\section{Illustrative Examples} 
In this section we apply the Analyticity approach to the following 
classical equations of the Mathematical Physics, the Schr\"odinger, the 
heat and the linear Korteweg-de Vries (KdV) equations:
\beq\label{eq:ex1}
{\partial u\over \partial t}-\alpha {\partial^2 u\over \partial x^2}=f,
~~~\alpha =i,1~~~~x\in V,~t>0,
\eeq
\beq\label{eq:ex2}
{\partial u\over \partial t}-\eta{\partial^3 u\over \partial x^3}=f,
~~\eta =\pm 1,~~x\in V,~t>0,
\eeq
prototype examples respectively of second and third order 
evolutionary PDEs and basic universal models for the description of 
dispersive and diffusive phenomena, where the space domain $V$ is either the segment 
$(0,L)$ or the semiline $(0,\infty )$. Hereafter the BVs will be 
indicated by
\beq\label{eq:BVs}
v^{(j)}_0(t):={\partial^ju\over \partial x^j}(x,t)|_{x=0},~~~
v^{(j)}_L(t):={\partial^ju\over \partial x^j}(x,t)|_{x=L},~~~j\in{\cal N}
\eeq
and their Fourier transforms by $\hat v^{(j)}_0(q),~\hat v^{(j)}_L(q)$:
\beq\label{eq:FTBVs}
\hat v^{(j)}_0(q):=\int\limits_0^\infty dte^{-iqt}v^{(j)}_0(t),~~~
\hat v^{(j)}_L(q):=\int\limits_0^\infty dte^{-iqt}v^{(j)}_L(t).
\eeq

We also apply the method to the study of IBV problems for the 
multimensional analogue of equation 
(\ref{eq:ex1}), for $\alpha =i$:
\beq\label{eq:ex3}
{\partial u\over \partial t}-i\bigtriangleup u=f , ~~~x\in V,~t>0,~~~
\bigtriangleup :=\bigtriangledown\cdot\bigtriangledown =
\sum\limits_{j=1}^n{\partial^2\over\partial {x_j}^2}
\eeq
in the $n$-dimensional Quadrant
\beq\label{eq:quadrantn}
V=\{{\bf x}:~x_j\ge 0,~j=1,..,n\}. 
\eeq
The corresponding BVs will be indicated by:
\bea\label{eq:BVn}
\ba{c}
v^{(0)}_{0j}({\bf x}_j,t)=u({\bf x},t)|_{x_j=0},~~
v^{(1)}_{0j}({\bf x}_j,t)={\partial u\over \partial x_j}({\bf x},t)|_{x_j=0}
\ea
\eea 
and their FTs by:
\bea\label{eq:FTBVn}
\ba{c}
\hat v^{(m)}_{0j}({\bf k}_j,q)=
\int\limits_0^{\infty}dt\int\limits_{V_j}d{\bf x}_j
e^{-i({\bf k}_j\cdot{\bf x}_j+qt)}v^{(m)}_{0j}({\bf x}_j,t),~~~m=0,1.
\ea
\eea
In equations (\ref{eq:BVn})-(\ref{eq:FTBVn}) 
${\bf x}_j=(x_1,..,\check{x_j},..,x_n)\in {\cal R}^{n-1}$, 
${\bf k}_j=(k_1,..,\check{k_j},..,k_n)\in {\cal R}^{n-1}$,  
$\int_{V_j}d{\bf x}'_j=\int_0^{L_1}dx'_1\cdot\cdot(\check{\int_0^{L_j}dx'_j})
\cdot\cdot\int_0^{L_n}dx'_n$ and  
the superscript $\check{~}$ indicates that the quantity underneath is 
removed.

The application of the Analyticity approach to higher order problems 
and to other relevant examples will be presented in \cite{DMS2}. 

\subsection{The second order PDEs (\ref{eq:ex1})}

In this case, equations (\ref{eq:Gformula}) and (\ref{eq:choice1})  
imply: 
\bea\label{eq:J1}
\ba{c}
\tilde {\cal L}=-{\partial\over \partial t}-\alpha 
{\partial^2\over \partial x^2},
~~~~~~ J=(ab,\alpha [a{\partial b\over\partial x}-
b{\partial a\over\partial x}]), \\
{\cal L}(i{k},iq)=i(q-i\alpha k^2).
\ea
\eea
In addition, if $V$ is the segment $(0,L)$, equation (\ref{eq:FTu2}) yields 
\bea\label{eq:FTu1box}
\ba{c}
\hat u(k,q)={\hat{\cal N}(k,q)\over i(q-i\alpha k^2)}, \\
\hat{\cal N}(k,q)=\hat f(k,q)+\hat u_0(k)-\alpha \left(
[\hat v^{(1)}_0(q)+ik\hat v^{(0)}_0(q)]- 
e^{-ikL}[\hat v^{(1)}_L(q)+ik\hat v^{(0)}_L(q)]\right) 
\ea
\eea
and the Fourier representation (\ref{eq:Frepr2}) of the solution takes the 
following form:
\bea\label{eq:Frepr1box}
\ba{c}
u({x},t)=\int_{{\cal R}^2}{dqdk \over (2\pi )^2i}
e^{i(kx+qt)}{\hat f({k},q)\over q-i\alpha k^2-i0}
+\int_{{\cal R}}{d{k} \over 2\pi}
e^{i{k}{x}-\alpha k^2t}\hat u_0({k}) + \\
\int_{\partial {\cal K}^{(\alpha )}_1}{dk\over 2\pi i}e^{ikx-\alpha k^2t}
[\hat v^{(1)}_0(i\alpha k^2)+ik\hat v^{(0)}_0(i\alpha k^2)]+ \\
\int_{\partial {\cal K}^{(\alpha )}_0}{dk\over 2\pi i}e^{ik(x-L)-\alpha k^2t}
[\hat v^{(1)}_L(i\alpha k^2)+ik\hat v^{(0)}_L(i\alpha k^2)],~~x\in (0,L),~~t>0,
\ea
\eea
where ${\cal K}^{(i)}_1$ and ${\cal K}^{(i)}_0$ are respectively the first 
and third quadrant of the complex $k$ - plane, ${\cal K}^{(1)}_m=
\hat\rho_{\pi\over 4}{\cal K}^{(i)}_m,~m=0,1$, where $\hat\rho_{\pi\over 4}$ 
is the $\pi /4$ rotation operator: $\hat\rho_{\pi\over 4}:~
k\to~e^{i\pi\over 4}k$ and $\partial{\cal K}$ is the 
counterclockwise oriented boundary of ${\cal K}$ (see Figs 1a,b).

The corresponding expressions for the semiline or for the infinite line 
cases, with rapidly decreasing conditions at $\infty$,  
follow immediately from the ones above, setting 
$\hat v^{(0)}_L=\hat v^{(1)}_L=0$ in the semiline case, or setting 
$\hat v^{(0)}_0=\hat v^{(1)}_0=\hat v^{(0)}_L=\hat v^{(1)}_L=0$ in the 
infinite line case.  

It is instructive to first apply the Analyticity approach to the simplest 
case in which the space domain is the whole space, with rapidly 
decreasing BVs at $x=\pm\infty$. 

\subsubsection{The whole line $V=(-\infty ,\infty )$}

Equation (\ref{eq:FTu1box}b) reduces to
\beq\label{eq:N1line}
\hat {\cal N}(k,q)=\hat f({k},q)+\hat u_0(k)
\eeq
and the admissibility argument imposes that 
$\hat {\cal N}(k,q)/{\cal L}$ be defined in 
$(k,q)\in{\cal A}={\cal R}\otimes \bar{\cal I}_q$ and be 
analytic in 
$q\in{\cal I}_q$, $\forall {k}\in{\cal R}$.  If 
$\alpha =1$, the denominator is singular for $q=ik^2$, $k\in {\cal R}$, 
outside the definition domain, and no regularization is needed. If, instead, 
$\alpha =i$, the denominator is singular for $q=-k^2<0$, on the real negative 
axis, at the boundary of the analyticity domain, 
and the only analyticity constraint to be  
fulfilled is the Denominator Regularization (\ref{eq:reg}):
\beq\label{eq:reg1}
{\cal L}(ik,iq)=i(q-i\alpha k^2)\to {\cal L}_{reg}(ik,iq)=i(q-i\alpha k^2-i0).
\eeq

The regularization (\ref{eq:reg1}) is sufficient to make the RHS of 
(\ref{eq:FTu1box}a) an admissible $FT_{\cal D}$, from which we recover 
the well-known Fourier representation of the solution of equations  
(\ref{eq:ex1}): 
\beq
u({x},t)=
\int_{{\cal R}^{2}}{dqd{k}e^{i({k}{x}+qt)} \over (2\pi )^{2}i}
{\hat f({k},q)\over q-i\alpha k^2-i0}+ 
\int_{{\cal R}}{d{k} \over 2\pi }
e^{i{k}{x}-\alpha k^2t}\hat u_0(k),~~~x\in {\cal R},~~t>0.
\eeq
We now proceed considering semi - compact and compact domains.

\subsubsection{The semiline $V=(0,\infty )$}

In this case:
\beq\label{eq:N1semiline}
\hat{\cal N}(k,q)=\hat f(k,q)+\hat u_0(k)-\alpha 
[\hat v^{(1)}_0(q)+ik\hat v^{(0)}_0(q)]
\eeq
and admissibility imposes that 
$\hat{\cal N}/{\cal L}$ be defined in 
${\cal A}=\bar{\cal I}_k\otimes \bar{\cal I}_q$, be analytic in 
$q\in{\cal I}_q,~
\forall k\in\bar{\cal I}_k$ and be analytic in $k\in{\cal I}_k,~
\forall q\in\bar{\cal I}_q$. Therefore the singularity manifolds  
$\cal A\cap{\cal S}^{(\alpha )}$, corresponding to $\alpha =i,1$, are   
parametrizable either in terms of $k$ or in terms of $q$ in the 
following way:
\beq\label{eq:SM1semiline}
{\cal A}\cap{\cal S}^{(\alpha )}=\{q=i\alpha k^2,~k\in\overline{{\cal K}^{(\alpha )}_0}\}=
\{k=k^{(\alpha )}_0(q),~\pi \le arg~q\le 2\pi\},
\eeq
where 
\bea
k^{(\alpha )}_0(q)=\left\{
\ba{c}
iq^{1\over 2},~~\alpha =i \\
e^{{3\over 4}\pi i}q^{1\over 2},~~\alpha =1.
\ea 
\right.
\eea

If $\alpha =1$, there is no singularity on the real axis and no 
regularization is needed. If $\alpha =i$, 
there are two singularities for $k\in {\cal R}$; that corresponding 
to $k<0$ is cured by the ZRC (\ref{eq:ZRC}), while that 
corresponding to $k>0$ is cured instead by the 
regularization (\ref{eq:reg1}). 

The ZRC (\ref{eq:ZRC}) is conveniently parametrized in terms of $q$ 
in the following way:
\beq\label{eq:ZRC1semiline}
\hat{\cal N}(k^{(\alpha )}_0(q),q)=
\hat f(k^{(\alpha )}_0(q),q)+
\hat u_0(k^{(\alpha )}_0(q))-\alpha 
[\hat v^{(1)}_0(q)+ik^{(\alpha )}_0(q)\hat v^{(0)}_0(q)]=0,
\eeq
for $\pi \le arg~q\le 2\pi$.  
It is one equation involving $4$ $FT's$ which are therefore dependent. 
If we are 
interested in solving the Dirichelet and Neumann problems, we use  
this ZRC to express the unknown BVs in terms of the known 
ones:
\bea\label{eq:solZRC1semiline}
\ba{c}
Dirichelet ~problem:~~~~\alpha\hat v^{(1)}_0(q)=
\hat f(k^{(\alpha )}_0(q),q)+
\hat u_0(k^{(\alpha )}_0(q))-\alpha ik^{(\alpha )}_0(q) 
\hat v^{(0)}_0(q), \\
Neumann ~problem:~~~i\alpha k^{(\alpha )}_0(q)\hat v^{(0)}_0(q)=
\hat f(k^{(\alpha )}_0(q),q)+\hat u_0(k^{(\alpha )}_0(q))-\alpha 
\hat v^{(1)}_0(q),
\ea
\eea
for $\pi \le arg~q\le 2\pi$. It is easy to see from 
(\ref{eq:solZRC1semiline}) that the unknown BVs 
define admissible FTs which, inserted in (\ref{eq:Frepr1box}), give the 
wanted solution of the Dirichelet and Neumann problems. 

We remark that the ZRC (\ref{eq:ZRC1semiline}) could also be solved for 
$\hat u_0$ (using now, for convenience, the variable $k$):
\beq\label{eq:solZRC1u0}
\hat u_0(k)=-\hat f(k,i\alpha k^2)+\alpha 
[\hat v^{(1)}_0(i\alpha k^2)+ik\hat v^{(0)}_0(i\alpha k^2)],~~~
k\in \overline{{\cal K}^{(\alpha )}_0}
\eeq
but, in this case, the solution would not be, in general, an 
admissible FT, since the RHS of (\ref{eq:solZRC1u0}) cannot be extended 
to the rest of the lower half $k$ - plane. Even in the special case in which the 
forcing and the assigned BVs were on a compact support in $t$, 
corresponding to entire FTs, the solution $\hat u_0(k)$ would not be   
admissible, because it would not possess, in general, the proper 
asymptotics. This means 
that the (unphysical) problem in which we 
assign arbitrarely $u$ and its space derivative at $x=0$ cannot be treated 
by this method, unless the above BVs are suitably 
constrained. 

\subsubsection{The segment $V=(0,L)$}

Now admissibility implies that $\hat{\cal N}/{\cal L}$ be defined in 
${\cal A}={\cal C}\otimes \bar{\cal I}_q$, be analytic in $q\in{\cal I}_q,~
\forall k\in {\cal C}$ and be analytic in $k\in {\cal C},~
\forall q\in\bar{\cal I}_q$, with proper asymptotics for large $|k|$ and/or 
$|q|$ in the analyticity 
regions. Therefore the singularity manifolds on which 
the ZRC (\ref{eq:ZRC}) is defined are now the unions of two sectors:
\bea\label{eq:SM1box}
\ba{c} 
{\cal A}\cap{\cal S}^{(\alpha )}=
\cup_{m=0}^1\{q=i\alpha k^2,~k\in\overline{{\cal K}^{(\alpha )}_m}\}= \\
\cup_{m=0}^1\{k=k^{(\alpha )}_m(q)=(-)^{m}k^{(\alpha )}_0(q),~~
\pi \le arg~q\le 2\pi\}.
\ea
\eea

\begin{center}
\mbox{\epsfxsize=6cm \epsffile{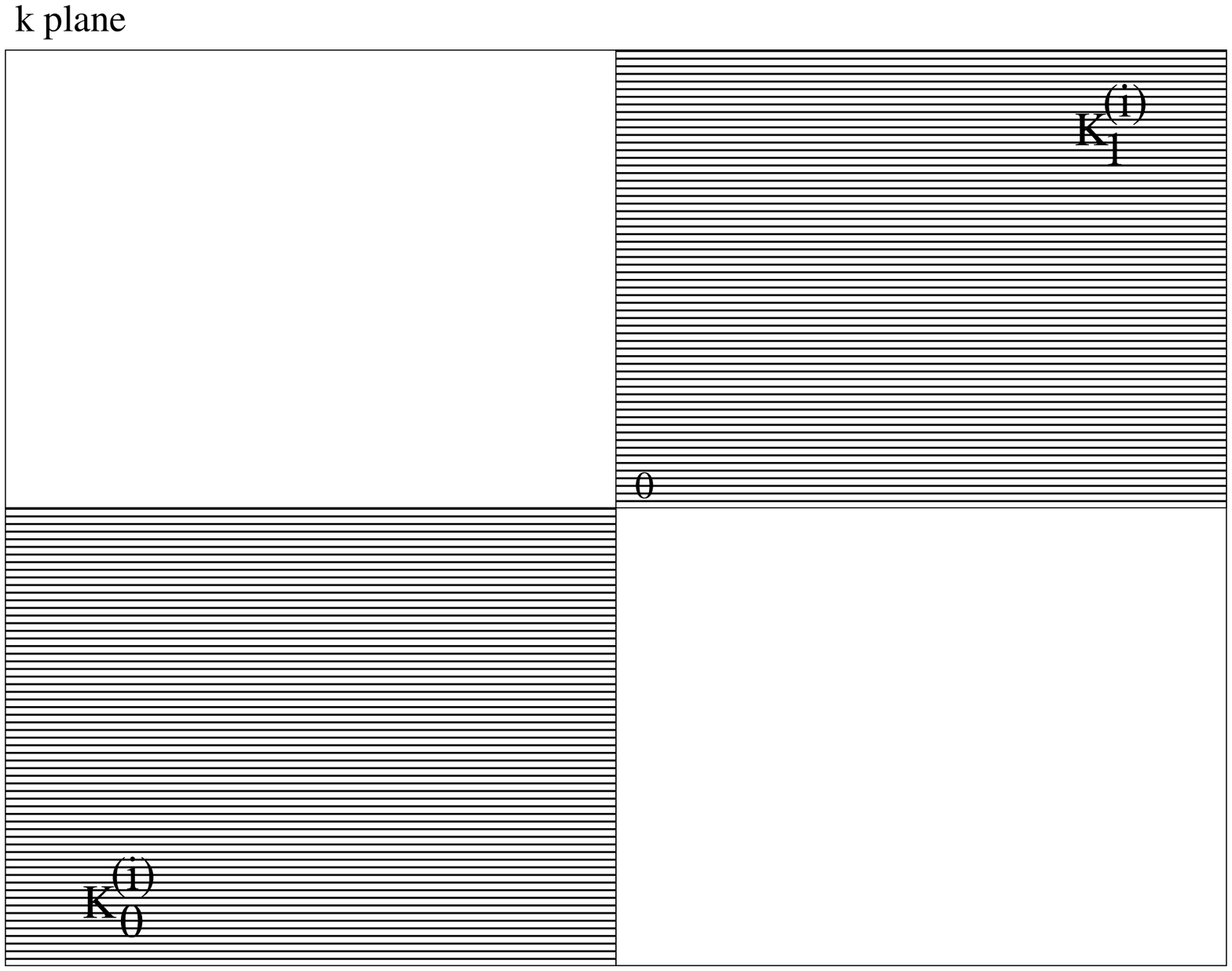}~~
\epsfxsize=6cm \epsffile{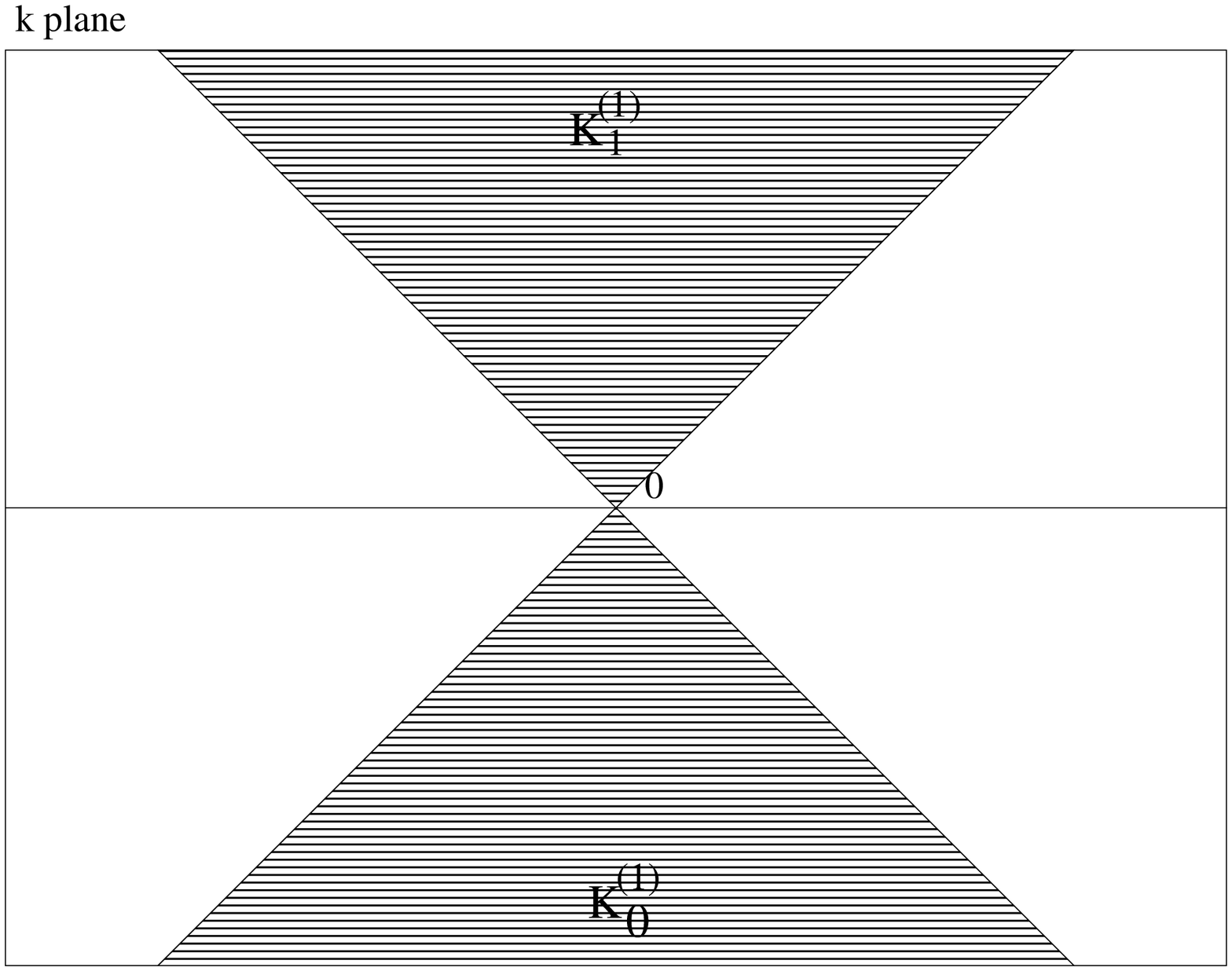}}
\end{center}

\noindent
Fig.1a$~~~~~$The SM ${\cal A}\cap{\cal S}^{(\alpha )}$ ($\alpha =i$)$~~~~~~$
Fig.1b$~~~~~$The SM ${\cal A}\cap{\cal S}^{(\alpha )}$ ($\alpha =1)$

\vskip 5pt
Both singularities on the real axis are cured 
by the ZRC and no regularization is needed. The  
regularization (\ref{eq:reg1}), however, is still  
introduced, according to the Remark 2 of \S 2.4, 
in computing the Fourier representation (\ref{eq:Freprboxn}) 
of the solution. The ZRC (\ref{eq:ZRC}), conveniently parametrized using  
$q$, consists of the following system of two linear algebraic equations:
\beq\label{eq:ZRC1boxbis}
\hat{\cal N}(k^{(\alpha )}_m(q),q)=0,~~~~~m=0,1~~~~~
\pi\le arg~q\le 2\pi
\eeq
containing four BVs. Therefore we expect to be allowed to assign 
arbitrarely two out of four BVs. To establish which pairs of 
BVs can be assigned arbitrarily, one should impose that the 
corresponding solutions of the algebraic system define admissible 
FTs; i.e., the following two conditions must be satisfied. 

\noindent
i) The system must be  uniquely solvable for the unknown pair of BVs  
in its definition domain. More precisely, indicating by $M$ the $2\times 2$ 
matrix of the coefficients of the unknown BVs, the admissibility 
condition imposes that the countable set $\{q_j\}_{j\in {\cal N}}$  
of zeroes of $\det M$, {\it the spectrum of the IBV 
problem}, lie outside the analyticity domain:
\beq
q_j\notin {\cal I}_q,~~~j\in{\cal N}. 
\eeq

\noindent
{\bf ii)} The solution of the system must define admissible Fourier 
Transforms; in particular, it must exhibit the proper asymptotics in the 
analyticity domain.

Studying first the asymptotics of (\ref{eq:ZRC1boxbis}), one infers without 
any effort which pairs of BVs {\bf cannot} be assigned arbitrarily. The 
asymptotics 
of (\ref{eq:ZRC1boxbis}) imply immediately that the following expressions:
\bea\label{eq:ZRC1asym}
\ba{c}
\hat f(k^{(\alpha )}_0(q),q)+\hat u_0(k^{(\alpha )}_0(q))-\alpha 
[\hat v^{(1)}_0(q)+ik^{(\alpha )}_0(q)\hat v^{(0)}_0(q)], \\ 
e^{ik^{(\alpha )}_0(q)L}[\hat f(-k^{(\alpha )}_0(q),q)+\hat u_0(-k^{(\alpha )}_0(q))]+
\alpha [\hat v^{(1)}_L(q)-ik^{(\alpha )}_0(q)\hat v^{(0)}_L(q)]
\ea
\eea
are exponentially small for $q\sim\infty$ in $\pi \le arg~q\le 2\pi$. 
Since the asymptotic series of the 
admissible FTs appearing in the LHS of equations (\ref{eq:ZRC1asym}) are 
inverse 
power series of $q^{1\over 2}$, equations (\ref{eq:ZRC1asym}) impose severe 
constraints on the involved functions, implying  that:

\noindent
asymptotic admissibility is compatible  with assigning at $x=0$ any BV 
between $(v^{(0)}_0,v^{(1)}_0)$ and, at $x=L$, any BV between 
$(v^{(0)}_L,v^{(1)}_L)$. It is not compatible instead with assigning  
arbitrarily the pairs $(v^{(1)}_0,v^{(0)}_0)$ or  
$(v^{(1)}_L,v^{(0)}_L)$.

To complete our analysis, we must check if the spectrum associated with the 
IBV problems compatible with the asymptotics lie outside the definition 
domain. The analysis is straightforward and produces the following 
results.

\noindent
{\bf Proposition (the spectrum)}. {\it Assigning arbitrarely 
$(v^{(0)}_0,v^{(0)}_L)$ 
(the Dirichelet problem) or $(v^{(1)}_0,v^{(1)}_L)$ (the Neumann problem), 
the spectrum is characterized by the equation $\sin (kL)=0~\Leftrightarrow~ 
k_m={\pi m\over L},~m\in{\cal Z}$ and is  
given by the negative eigenvalues $\{q_m\}_{n\in{\cal N}},~
q_m=-k^2_m=-({\pi m\over L})^2,~m\in{\cal N}$, if $\alpha =i$, and by the 
purely imaginary eigenvalues  
$q_m=ik^2_m=i({\pi m\over L})^2,~m\in{\cal N}$, if $\alpha =1$. 
Assigning instead $(v^{(0)}_0,v^{(1)}_L)$ or 
$(v^{(1)}_0,v^{(0)}_L)$ (the mixed problems), the spectrum is characterized 
by the equation $\cos (kL)=0~\Leftrightarrow~ 
k_m={\pi\over L}(2m+1),~m\in{\cal Z}$ and is given by the negative 
eigenvalues $\{q_m\}_{m\in{\cal N}},~q_m=-k^2_m=-({\pi\over L})^2(2m+1)^2,
~m\in{\cal N}$, if $\alpha =i$, and by the 
purely imaginary eigenvalues  
$q_m=ik^2_m=i({\pi \over L})^2(2m+1)^2,~m\in{\cal N}$, if $\alpha =1$}.

For $\alpha =1$ the spectrum lies outside the analyticity 
region and the solutions of the algebraic system (\ref{eq:ZRC1boxbis}) 
define directly admissible FTs; if $\alpha =i$ the solutions of the algebraic 
system (\ref{eq:ZRC1boxbis}) define admissible FTs 
after moving these singularities a bit off the real $q$ - 
axis, outside the definition domain (again a regularization!). We conclude 
that all the IBV problems compatible with admissible asymptotics turn out to 
be well-posed:

\vskip 5pt
\noindent
{\bf IBV problems for the Schr\"odinger and heat equations (\ref{eq:ex1}) 
 are well-posed assigning 
at $x=0$ any BV among $(v^{(0)}_0,v^{(1)}_0)$ and at $x=L$ any BV among 
$(v^{(0)}_L,v^{(1)}_L)$}.
\vskip 5pt

It is interesting to remark that, if one insisted, instead, in solving an IBV 
problem in which the BVs $(\hat v^{(1)}_0(q),~\hat v^{(0)}_0(q))$ are 
assigned, the correspondig algebraic system would be always uniquely solvable 
(no point spectrum would arise), but the solution would exhibit an 
exponential blow up at $q\sim\infty$ in the analyticity region, that cannot  
be accepted. This undesired blow up could be cured 
if the assigned 
BVs were related by the (additional) analyticity constraint:
\beq\label{eq:addAC}
\hat f(k^{(\alpha )}_0(q),q)+\hat u_0(k^{(\alpha )}_0(q))-\alpha 
[\hat v^{(1)}_0(q)+ik^{(\alpha )}_0(q)\hat v^{(0)}_0(q)]=0,
\eeq
implying the following admissible solutions of the algebraic system 
(\ref{eq:ZRC1boxbis}):
\bea
\ba{c}
\hat v^{(1)}_L(q)=-ik^{(\alpha )}_0(q)\hat v^{(0)}_L(q)=-{e^{-ik^{(\alpha )}_0(q)L}\over 2\alpha }
[\hat f(k^{(\alpha )}_0(q),q)+\hat f(-k^{(\alpha )}_0(q),q)+ \\
\hat u_0(k^{(\alpha )}_0(q))+\hat u_0(-k^{(\alpha )}_0(q))
-2\alpha \hat v^{(1)}_0(q)].
\ea
\eea
The additional analyticity contraint (\ref{eq:addAC}) 
is not surprising at all, since it is 
nothing but the ZRC of the semiline problem. Similarly, 
assigning the right boundary conditions ($v^{(0)}_L,~v^{(1)}_L$), the unknowns 
$\hat v^{(0)}_0$ and $\hat v^{(1)}_0$ would exhibit again an exponential 
blow up which 
cannot be accepted; an admissible asymptotics would be  guaranteed now by the 
(additional) analyticity constraint:
\beq
e^{-ik^{(\alpha )}_0(q)L}[\hat f(-k^{(\alpha )}_0(q),q)+\hat u_0(-k^{(\alpha )}_0(q))]+
\alpha [\hat v^{(1)}_L(q)-ik^{(\alpha )}_0(q)\hat v^{(0)}_L(q)]=0, 
\eeq
implying the following solution of the algebraic system:
\bea
\ba{c}
\hat v^{(1)}_0(q)=ik^{(\alpha )}_0(q)\hat v^{(0)}_0(q)={1\over 2\alpha }
[\hat f(k^{(\alpha )}_0(q),q)+\hat f(-k^{(\alpha )}_0(q),q)+ \\
\hat u_0(k^{(\alpha )}_0(q))+\hat u_0(-k^{(\alpha )}_0(q))
+2\alpha e^{-ik^{(\alpha )}_0(q)L}\hat v^{(1)}_L(q)].
\ea
\eea

\subsubsection{The periodic problem}

If we assume $L$-periodicity of $u$ and $u_x$, then $\hat v^{(1)}_0=\hat v^{(1)}_L=:\hat v^{(1)}$,  
$\hat v^{(0)}_0=\hat v^{(0)}_L=:\hat v^{(0)}$ and the algebraic system (\ref{eq:ZRC1boxbis}) 
consists now of two equations for two BVs, which have to be treated therefore 
as unknowns. The solutions of this system read:
\bea
\ba{c}
\hat v^{(1)}(q)={1\over 2\alpha }\left( 
{\hat f(k^{(\alpha )}_0(q),q)+\hat u_0(k^{(\alpha )}_0(q))\over 1-e^{-ik^{(\alpha )}_0(q)L}}+
{\hat f(-k^{(\alpha )}_0(q),q)+\hat u_0(-k^{(\alpha )}_0(q))\over 1-e^{ik^{(\alpha )}_0(q)L}}
\right), \\
\hat v^{(0)}(q)={1\over 2i\alpha k^{(\alpha )}_0(q)}\left( 
{\hat f(k^{(\alpha )}_0(q),q)+\hat u_0(k^{(\alpha )}_0(q))\over 1-e^{-ik^{(\alpha )}_0(q)L}}-
{\hat f(-k^{(\alpha )}_0(q),q)+\hat u_0(-k^{(\alpha )}_0(q))\over 1-e^{ik^{(\alpha )}_0(q)L}}
\right).
\ea
\eea
They satisfy asymptotic admissibility and the spectrum, characterized by 
the equation $1-e^{\pm ikL}=0$, ($\Rightarrow~k_n={2\pi\over L}n,~n\in{\cal Z}$), 
is given by $q_n=-k^2_n=-({2\pi\over L})^2n^2,~n\in{\cal N}$, for $\alpha =i$, and by 
$q_n=ik^2_n=i({2\pi\over L})^2n^2,~n\in{\cal N}$, for $\alpha =1$; therefore the usual 
regularization is needed again in the Schr\"odinger case. We conclude that 

\noindent
{\bf the 
periodic problem for equations (\ref{eq:ex1}), in which one imposes the $L$-
periodicity of $u$ and $u_x$, is well-posed and no BV can be assigned arbitrarely}.
  
\vskip 10pt 
\noindent
{\bf Remark} We remark that the Fourier transforms of the 
unknown boundary functions exhibit generically a branch point at $q=0$,  
due to the well-known slow decay as $t\to\infty$ of the solutions of 
the dispersive evolution equation under investigation. 

The above procedure generalizes with no difficulties to higher order problems. 
In the following we concentrate on a third order problem only.

\subsection{The linear KdV equation}
In this section we investigate IBV problems for $3^{rd}$ order operators,  
illustrating the method on the simplest possible example (\ref{eq:ex2}).

Since the group velocity $v_g=3\eta k^2$ 
of the associated wave packet is positive (negative ) 
for $\eta$ positive (negative), we have the following expectations. 
In the semiline case, one should be able to 
assign at $x=0$ more BVs for positive $\eta$ than for negative $\eta$. In the 
segment case, 
for $\eta$ positive one can assign arbitrarely more BVs at $x=0$ than at 
$x=L$ (and viceversa 
for $\eta$ negative). The precise  indication of ``how many'' and ``which'' 
BVs can be assigned in order to have a well-posed IBV problem 
follows again in a straightforward way from the Analyticity approach.

Equations (\ref{eq:Gformula}) and (\ref{eq:choice1})  
imply: 
\bea\label{eq:J2}
\ba{c}
\tilde {\cal L}=-{\cal L},
~~~~~~ J=(ab,-\eta [b{\partial^2 a\over\partial x^2}-
{\partial b\over\partial x}{\partial a\over\partial x}+
{\partial^2 b\over\partial x^2}a]), \\
{\cal L}(i{\bf k},iq)=i(q+\eta k^3).
\ea
\eea
In addition, if $V$ is the segment $(0,L)$, equation (\ref{eq:FTu2}) yields 
\bea\label{eq:FTu2box}
\ba{c}
\hat u(k,q)=-i{\hat{\cal N}(k,q)\over q+\eta k^3}, \\
\hat{\cal N}(k,q)=\hat f(k,q)+\hat u_0(k)-\eta \left(
[\hat v^{(2)}_0(q)+ik\hat v^{(1)}_0(q)-k^2\hat v^{(0)}_0(q)]- \right. \\ 
\left. e^{-ikL}[\hat v^{(2)}_L(q)+ik\hat v^{(1)}_L(q)-k^2\hat v^{(0)}_L(q)]\right)
\ea
\eea
and the Fourier representation (\ref{eq:Frepr2}) of the solution takes the 
following form:
\bea\label{eq:Frepr2box}
\ba{c}
u({x},t)=\int_{{\cal R}^2}{dqdk \over (2\pi )^2i}
e^{i(kx+qt)}{\hat f({k},q)\over q+\eta k^3-i0}
+\int_{{\cal R}}{d{k} \over 2\pi}
e^{i({k}{x}-\eta k^3t)}\hat u_0({k}) - \\
\eta (
\int_{\gamma^{(\eta )}_{0}}{dk\over 2\pi}e^{i(kx-\eta k^3t)}
[\hat v^{(2)}_0(-\eta k^3)+ik\hat v^{(1)}_0(-\eta k^3)-
k^2\hat v^{(0)}_0(-\eta k^3)]+  \\
\int_{\gamma^{(\eta )}_{L}}{dk\over 2\pi}e^{i[k(x-L)-\eta k^3t]}
[\hat v^{(2)}_L(-\eta k^3)+ik\hat v^{(1)}_L(-\eta k^3)-
k^2\hat v^{(0)}_L(-\eta k^3)]),~x\in (0,L),~t>0,
\ea
\eea
where $\gamma^-_{0}=\partial{\cal K}^{(-)}_0$, 
$\gamma^-_{L}=\partial{\cal K}^{(-)}_1\cup\partial{\cal K}^{(-)}_2$, 
$\gamma^+_{0}=\partial{\cal K}^{(+)}_1\cup\partial{\cal K}^{(+)}_2$, 
$\gamma^+_{L}=\partial{\cal K}^{(+)}_0$, 
\beq
{\cal K}^{(-)}_m=\{k:~{\pi\over 3}(2m+1)\le~arg~k\le{\pi\over 3}(2m+2) \},~~
{\cal K}^{(+)}_m=\hat\rho_{\pi} {\cal K}^{(-)}_m,~~m=0,1,2, 
\eeq
and $\hat\rho_{\pi}$ is the involution $\hat\rho_{\pi} :~k\to -k$.

\noindent
\subsubsection{The segment $V=(0,L)$}

Now  
$\hat{\cal N}/{\cal L}$ must be defined in 
${\cal A}={\cal C}\otimes \bar{\cal I}_q$, analytic in $q\in{\cal I}_q,~
\forall k\in {\cal C}$ and analytic in $k\in {\cal C},~
\forall q\in\bar{\cal I}_q$, with proper asymptotics for large $|k|$ and/or 
$|q|$ in the analyticity regions. Therefore the 
singularity manifolds ${\cal A}\cap{\cal S}^{(\eta )}$, corresponding to 
$\eta =\pm 1$, are given by (see Figs 2a,b):
\beq
{\cal A}\cap{\cal S}^{(\eta )}=\cup_{m=0}^2\{q=-\eta k^3,~
k\in {\cal K}^{(\eta )}_m \}=
\cup_{m=0}^2\{k=k^{(\eta )}_m(q),~~\pi\le~arg~q\le 2\pi \}
\eeq
where $k^{(\eta )}_m(q)=-\eta\rho_m q^{1\over 3}$ and 
$\rho_m,~m=0,1,2$ are the $3$ roots of unity:
\beq
\rho_m=e^{{2\pi i\over 3}m},~m=0,1,2.
\eeq

\begin{center}
\mbox{\epsfxsize=6cm \epsffile{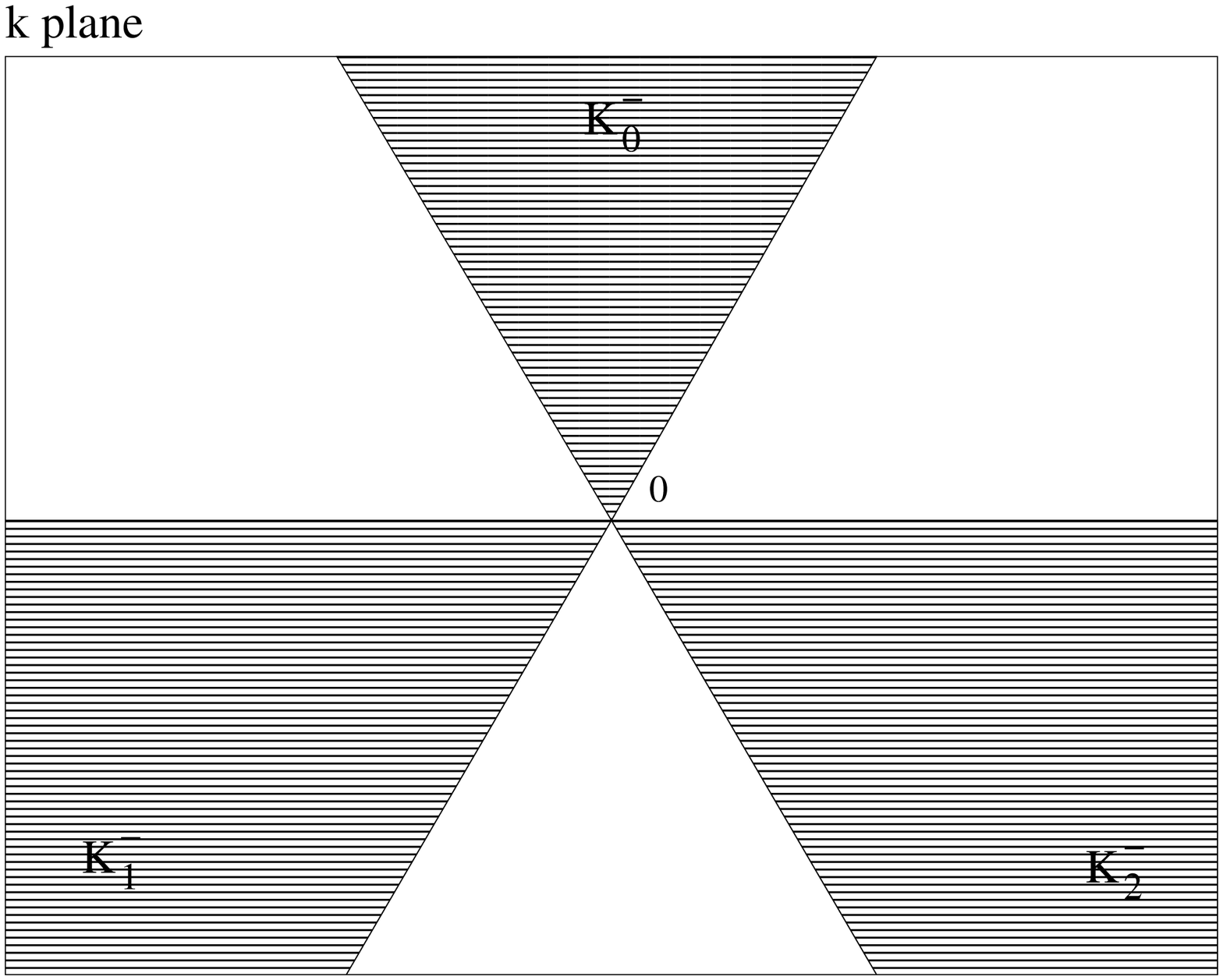}~~
\epsfxsize=6cm \epsffile{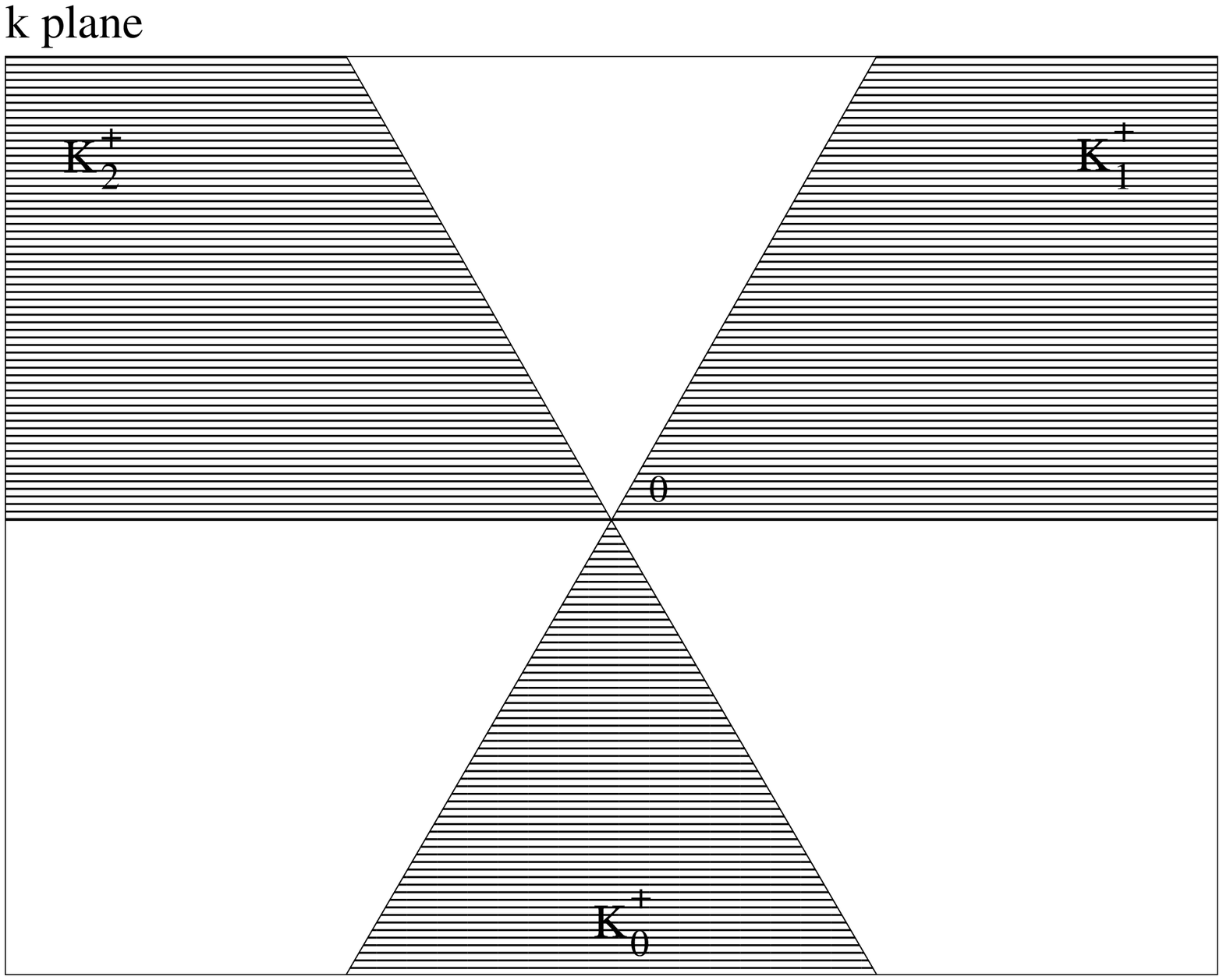}}
\end{center}

\noindent
Fig.2a$~~~~~$The SM ${\cal A}\cap{\cal S}^{(\eta )}$ ($\eta =-1$)$~~~~~~$
Fig.2b$~~~~~$The SM ${\cal A}\cap{\cal S}^{(\eta )}$ ($\eta =1)$

\vskip 5pt
\noindent
The ZRC (\ref{eq:ZRC}) consists of the following three equations:
\beq\label{eq:ZRC2box}
\hat{\cal N}(k^{(\eta )}_m(q),q)=0,~~~
m=0,1,2,~~~\pi \le arg~q\le 2\pi .
\eeq
For $q\in{\cal R}$ there is one singularity on the real $k$ - axis, which 
is cured by one of the three equations (\ref{eq:ZRC2box})  
and no denominator regularization is then needed. The  
regularization (\ref{eq:reg1}), however, is still  
introduced, according to the Remark 2 of \S 2.2.2, in writing  
the Fourier representation (\ref{eq:Frepr2box}) 
of the solution. 

The 3 algebraic equations (\ref{eq:ZRC2box}) 
contain 6 BVs; therefore we expect to be allowed to assign 
independently only 3 BVs. As before, a quick asymptotic extimate selects the 
sets of 3 BVs which can be assigned independently, compatibly with  
asymptotic admissibility.  
The asymptotics of equations (\ref{eq:ZRC2box}) imply that 
the following expressions, respectively, for $\eta =-1$:
\bea\label{eq:asym-}
\ba{c}
e^{iq^{1\over 3}L}[\hat f(q^{1\over 3},q)+\hat u_0(q^{1\over 3})] -
[\hat v^{(2)}_L(q)+iq^{1\over 3}\hat v^{(1)}_L(q)-
q^{2\over 3}\hat v^{(0)}_L(q)], \\
\hat f(\rho_1 q^{1\over 3},q)+\hat u_0(\rho_1 q^{1\over 3}) +
[\hat v^{(2)}_0(q)+i\rho_1q^{1\over 3}\hat v^{(1)}_0(q)-
\rho_2q^{2\over 3}\hat v^{(0)}_0(q)], \\
\hat f(\rho_2 q^{1\over 3},q)+\hat u_0(\rho_2 q^{1\over 3}) +
[\hat v^{(2)}_0(q)+i\rho_2q^{1\over 3}\hat v^{(1)}_0(q)-
\rho_1q^{2\over 3}\hat v^{(0)}_0(q)],
\ea
\eea  
and for $\eta =1$: 
\bea\label{eq:asym+}
\ba{c}
\hat f(-q^{1\over 3},q)+\hat u_0(-q^{1\over 3}) -
[\hat v^{(2)}_0(q)-iq^{1\over 3}\hat v^{(1)}_0(q)-
q^{2\over 3}\hat v^{(0)}_0(q)], \\
e^{-i\rho_1q^{1\over 3}L}[\hat f(-\rho_1q^{1\over 3},q)+
\hat u_0(-\rho_1q^{1\over 3})] +
[\hat v^{(2)}_L(q)-i\rho_1q^{1\over 3}\hat v^{(1)}_L(q)-
\rho_2q^{2\over 3}\hat v^{(0)}_L(q)], \\
e^{-i\rho_2q^{1\over 3}L}[\hat f(-\rho_2q^{1\over 3},q)+
\hat u_0(-\rho_2q^{1\over 3})] +
[\hat v^{(2)}_L(q)-i\rho_2q^{1\over 3}\hat v^{(1)}_L(q)-
\rho_1q^{2\over 3}\hat v^{(0)}_L(q)],
\ea
\eea
are exponentially small for $q\sim\infty$ in $\pi \le arg~q\le 2\pi$. 
Therefore, reasoning as before, we see that: 

\noindent
i) for $\eta =-1$, a necessary and sufficient condition to 
obtain FTs with admissible asymptotics is to assign at $x=0$ any BV 
among $v^{(0)}_0,v^{(1)}_0,v^{(2)}_0$ (consequence of  
equations (\ref{eq:asym-}b,c)) and, at $x=L$, any two BVs among 
$v^{(0)}_L,v^{(1)}_L,v^{(2)}_L$ (consequence of  
equation (\ref{eq:asym-}a)); 

\noindent
ii) for $\eta =1$, a necessary and sufficient condition to 
obtain FTs with admissible asymptotics is to assign at $x=0$ any two BVs 
among $v^{(0)}_0,v^{(1)}_0,v^{(2)}_0$ (consequence of  
equation (\ref{eq:asym+}a)) and, at $x=L$, any BV among 
$v^{(0)}_L,v^{(1)}_L,v^{(2)}_L$ (consequence of 
equations (\ref{eq:asym+}a,b)).

Again, to complete our investigation, we must check if the spectrum 
associated with the above IBV problems selected by the asymptotic 
admissibility, lie entirely outside the analyticity domain of an 
admissible FT. 
It is easy to prove that it is indeed the case.

\noindent
{\bf Proposition (the spectrum of the IBV problem)} {\it Consider any IBV 
problem on the segment for equation (\ref{eq:ex2}) compatible with 
the asymptotic admissibility established above; i.e., in which, 
for $\eta =-1$, one assigns arbitrarely at $x=0$
any BV among $v^{(0)}_0,v^{(1)}_0,v^{(2)}_0$  and any two BVs at $x=L$
among $v^{(0)}_L,v^{(1)}_L,v^{(2)}_L$, and in which, for $\eta =1$, one 
assigns arbitrarely at $x=0$
any two BVs among $v^{(0)}_0,v^{(1)}_0,v^{(2)}_0$  and any BV at $x=L$
among $v^{(0)}_L,v^{(1)}_L,v^{(2)}_L$.    
For $\eta =-1$, let $v^{(n)}_0$ be the given 
BV at $x=0$ and $v^{(m)}_L$ be the unknown BV at $x=L$ while, for 
$\eta =1$, let $v^{(n)}_0$ be the unknown  
BV at $x=0$ and $v^{(m)}_L$ be the given BV at $x=L$. 
Then the corresponding spectrum is characterized by the following equation:
\beq
\Delta^{(\eta (m-n))}(k)=0,
\eeq    
where:}
\beq
\Delta^{(j)}(k):=e^{-ikL}+\rho^j_1e^{-\rho_1ikL}+
\rho^j_2e^{-\rho_2ikL}.
\eeq
The proof is tedious but straightforward and makes essential use of the 
well-known algebra of the roots of unity, which implies also that all 
the above IBV problems lead only to three (similar) purely 
imaginary discrete spectra $\{k^{(j)}_n\}_{n\in{\cal N}}$,  
characterized by the three equations 
$\Delta^{(j)}(k)=0,~j=0,1,2$. More precisely: 1) the 
spectrum characterized by equation $\Delta^{(0)}(k)=0$ is given by:
\bea
\ba{c}
k^{(0)}_n=-i(\zeta^{(0)}_n/L),~n\in{\cal N}^+:~~~
\{\zeta^{(0)}_n\}_{n\in\cal N^+}:~~~{e^{-{3\over 2}\zeta^{(0)}_n}\over 2}=
-\cos ({\sqrt{3}\over 2}\zeta^{(0)}_n ), \\
\Rightarrow~\zeta^{(0)}_n\sim {\pi\over \sqrt{3}}(2n-1),~n\ge 1.
\ea
\eea
2) The spectrum 
characterized by equation $\Delta^{(1)}(\zeta )=0$ is:
\bea
\ba{c}
k^{(1)}_n=-i(\zeta^{(1)}_n/L),~n\in{\cal N}:~~~
\{\zeta^{(1)}_n\}_{n\in\cal N}:~~~{e^{-{3\over 2}\zeta^{(1)}_n}\over 2}=
\cos ({\sqrt{3}\over 2}\zeta^{(1)}_n+{\pi\over 3} ), \\
\Rightarrow~\zeta^{(1)}_0=0,~~\zeta^{(1)}_n\sim {2\pi\over \sqrt{3}}(n-{5\over 6}),~n\ge 2.
\ea
\eea
3) The spectrum 
characterized by equation $\Delta^{(2)}(\zeta )=0$ is:
\bea
\ba{c}
k^{(2)}_n=-i(\zeta^{(2)}_n/L),~n\in{\cal N}:~~
\{\zeta^{(2)}_n\}_{n\in\cal N}:~~~{e^{-{3\over 2}\zeta^{(2)}_n}\over 2}=
\cos ({\sqrt{3}\over 2}\zeta^{(2)}_n-{\pi\over 3} ), \\
\zeta^{(2)}_0=0,~~\zeta^{(2)}_n\sim {2\pi\over \sqrt{3}}(n-{1\over 6}),~n\ge 1.
\ea
\eea
We conclude that all the three discrete spectra 
\beq
\{q^{(j)}_n\}_{n\in\cal N},~~~~q^{(j)}_n={k^{(j)}_n}^3=
i\left({\zeta^{(j)}_n\over L}\right)^3,~~~~j=0,1,2,
\eeq
associated with the above IBV problems 
lie on the positive imaginary axis of the complex $q$ plane, outside the 
analyticity domain of an admissible FT. Therefore:
\vskip 5pt
\noindent
{\bf 
IBV problems for equation (\ref{eq:ex2}) on the segment $(0,L)$ are 
well-posed iff: 

\noindent
i) for $\eta =-1$,  one assigns at $x=0$ any BV among 
$v^{(0)}_0,v^{(1)}_0,v^{(2)}_0$  and at $x=L$ any two BVs 
among $v^{(0)}_L,v^{(1)}_L,v^{(2)}_L$; 

\noindent
ii) for $\eta =1$, one 
assigns at $x=0$
any two BVs among $v^{(0)}_0,v^{(1)}_0,v^{(2)}_0$  and any BV at $x=L$
among $v^{(0)}_L,v^{(1)}_L,v^{(2)}_L$}.

\subsubsection{The periodic problem}

If we assume $L$-periodicity of $u$, $u_x$ and $u_{xx}$, then 
$v^{(j)}_0=v^{(j)}_L,~j=0,1,2$, the algebraic system (\ref{eq:ZRC2box}) consists 
now of three equations for three BVs, which have to be treated then as 
unknowns. The solution of this system satisfy asymptotic admissibility 
and the spectrum, characterized by the equations 
$1-e^{-i\rho_jk}=0,~j=0,1,2$ ($\Rightarrow~k_n={2\pi\over L}\rho^{-1}_jn,~
n\in{\cal Z},~j=0,1,2$), is given by the real numbers $q_n=-\eta k^3_n=
-\eta ({2\pi\over L})^3n^3,~n\in{\cal Z}$ and must be regularized in the usual way. 
We conclude that:

\noindent
{\bf the periodic problem for the linear KdV equation (\ref{eq:ex2}), 
in which one imposes $L$-periodicity to $u$, $u_x$ and $u_{xx}$, 
is well-posed and no BV can be assigned}. 

\subsubsection{The semiline $V=(0,\infty)$}

\vskip 5pt
\noindent
Taking the limit $L\to\infty$ of the results of \S 3.2.1 we immediately obtain  
the results on the semiline. In this case, the singularity 
manifolds are the restrictions of the above ones to the lower half $k$ plane. 
No spectrum arises and the asymptotic admissibility implies that:
\vskip 5pt
\noindent
{\bf 
IBV problems for equation (\ref{eq:ex2}) on the semiline $(0,\infty )$ are 
well-posed iff, for $\eta =-1$,  one assigns at $x=0$ any BV among 
$(v^{(0)}_0,v^{(1)}_0,v^{(2)}_0)$  and, for $\eta =1$, one 
assigns at $x=0$ any two BVs among $(v^{(0)}_0,v^{(1)}_0,v^{(2)}_0)$}.

\vskip 5pt
We remark that, in the cases treated so far, the spectra of all the IBV 
problems compatible with asymptotic admissibility lie always outside the 
analyticity domain. We do not have, however, a general argument  
excluding the situation in which part of the spectrum lie inside. Therefore 
the complete characterization of the spectrum, the only part of the method  
in which some technicality is involved, seems to be unavoidable and 
makes it difficult to prove general results for operators of arbitrary 
order.

The Analyticity approach applies nicely also to an arbitrary number of 
dimensions and next section is devoted to an illustration of it. The 
application of the method to higher order problems and to other relevant 
examples will be presented in \cite{DMS2}. 

\subsection{Multidimensional Schr\"odinger equation}

In this section we study the Dirichelet and 
Neumann problems for the Schr\"odinger equation 
(\ref{eq:ex3}) in the $n$-dimensional quadrant (\ref{eq:quadrantn}). Then:
\bea\label{eq:J3}
\ba{c}
\tilde {\cal L}=-{\partial\over \partial t}-i\bigtriangleup ,~~~~~~ 
J=(ab,i(a\bigtriangledown b-b\bigtriangledown a), \\
{\cal L}(i{\bf k},iq)=i(q+k^2),
\ea
\eea
where $k^2={\bf k}\cdot {\bf k}$. 
Equations (\ref{eq:FTu1}) and (\ref{eq:J3}) give the following 
expression of the Fourier transform of the solution 
in terms of the Fourier transforms of the 
forcing and of all the IBVs:
\bea\label{eq:FTuboxn}
\ba{c}
\hat u({\bf k},q)={\hat{\cal N}({\bf k},q)\over i(q+k^2)}, \\
\hat{\cal N}({\bf k},q):=\hat f({\bf k},q)+\hat u_0({\bf k})-i
\sum\limits_{j=1}^n [\hat v^{(1)}_{0j}({\bf k}_j,q)+
ik_j\hat v^{(0)}_{0j}({\bf k}_j,q)] .
\ea
\eea
The Fourier representation (\ref{eq:Frepr2}) of the solution reads:

\bea\label{eq:Freprboxn}
\ba{c}
u({\bf x},t)=
\int_{{\cal R}^{n+1}}{dqd{\bf k} \over (2\pi )^{n+1}i}
e^{i({\bf k}\cdot {\bf x}+qt)}
{\hat f({\bf k},q)\over q+k^2-i0}
+\int_{{\cal R}^n}{d{\bf k} \over (2\pi )^n}
e^{i({\bf k}\cdot {\bf x}-k^2t)}\hat u_0({\bf k})+  \\
\sum\limits_{j=1}^n\int\limits_{{\cal R}^{n-1}}{d{\bf k}_j\over 
(2\pi )^{n-1}}
\int_{\partial{\cal K}^{(i)}_1}{dk_j\over 2\pi i}\{ e^{i({\bf k}\cdot{\bf x}-k^2t)} 
[\hat v^{(1)}_{0j}({\bf k}_j,-k^2)+
ik_j\hat v^{(0)}_{0j}({\bf k}_j,-k^2)],
\ea
\eea
where $d{\bf k}_j=dk_1..\check{dk_j}..dk_n$.

In view of the distinguished parity properties of the Fourier transforms 
in (\ref{eq:FTuboxn}), we shall make an extensive use  of the parity 
operators: 
\beq\label{eq:parityproj}
\Delta_{\pm}=\prod\limits_{l=1}^n(1{\pm}\hat\sigma_l),~~~
\Delta^{(j)}_{\pm}=
\prod\limits_{l=1\atop l\ne j}^n(1{\pm}\hat\sigma_l),
\eeq
where $\hat\sigma_j$ is the involution $\hat\sigma_j:~k_j~\to~-k_j$.

In this multidimensional case, the FT of the solution is defined in ${\cal A}=
\bar{\cal I}_{k_1}\otimes \cdot\cdot\otimes \bar{\cal I}_{k_n}\otimes
\bar{\cal I}_{q}$, analytic for $q\in{\cal I}_{q}$, $\forall 
{\bf k}\in\bar{\cal I}_{k_1}\otimes\cdot\cdot\otimes{\bar{\cal I}}_{k_n}$ and 
in $k_j\in {\cal I}_{k_j}$, $\forall 
{\bf k}_j\in\bar{\cal I}_{k_1}\otimes \cdot\cdot\otimes 
\check{\bar{\cal I}_{k_j}}\otimes\cdot\cdot\otimes\bar{\cal I}_{k_n}$ and   
$\forall q\in\bar{\cal I}_{q}$. We found it convenient 
to study the ZRC in the $n$ different regions 
${\cal Q}^-_{j}\subset {\cal A}\cap{\cal S},~j=1,..,n$ 
defined by:
\bea\label{eq:defQj}
\ba{c}
{\cal Q}^-_{j}:=\{({\bf k},q)\in{\cal C}^{n+1}:~
 {\bf k}_j\in {\cal R}^{n-1},~\pi\le arg~q\le 2\pi ,
~k_j=\chi_j({\bf k}_j,q)\},
  \\
\chi_j({\bf k}_j,q):=i(q+{\bf k}_j\cdot{\bf k}_j)^{1\over 2}\in \overline{{\cal I}_k},
~~~~~~j=1,..,n.
\ea
\eea
Therefore the starting point of the analysis is the set of $n$ equations 
\beq\label{eq:ZRCquadrn}
\hat{\cal N}({\bf k},q)|_{k_j=\chi_j}=0,~~{\bf k}_j\in{\cal R}^{n-1},~~
\pi\le arg~q\le 2\pi ,~~~j=1,..,n.
\eeq

\noindent
{\it Dirichelet problem}. The parity properties in $k$ of the BV terms imply 
that the application of the parity operator $\Delta^{(j)}_-$ defined in 
(\ref{eq:parityproj}) to the $j^{th}$ equation (\ref{eq:ZRCquadrn}) 
eliminates all the $\hat v^{(1)}_{0}$s except $\hat v^{(1)}_{0j}$:
\bea\label{eq:elimjD}
\ba{c}
\Delta^{(j)}_-\hat v^{(1)}_{0j}({\bf k}_j,q)=
-\Delta^{(j)}_-(W({\bf k},q)|_{k_j=\chi_j}),~~~j=1,..,n,   \\
W({\bf k},q):=\hat f({\bf k},q)+i\hat u_0({\bf k})+
i\sum\limits_{j=i}^nk_j\hat v^{(0)}_{0j}({\bf k}_j,q)
\ea
\eea
and the analyticity properties of the $\hat v^{(1)}_{0}$s allow one to 
express them in terms of known quantities:
\bea\label{eq:solwquadrn}
\ba{c}
\hat v^{(1)}_{0j}({\bf k}_j,q)=
{\cal P}^{(j)}\Delta^{(j)}_-\hat v^{(1)}_{0j}({\bf k}_j,q)=
-{\cal P}^{(j)}\Delta^{(j)}_-(W({\bf k},q)|_{k_j=\kappa_j}),~~~j=1,..,n
\ea
\eea
applying the lower half plane analyticity projectors in all the 
$k$-variables (except $k_j$):
\beq\label{eq:analyticityproj}
{\cal P}^{(j)}=\prod\limits_{m=1\atop m\ne j}^n{\cal P}_m,~~~~~
{\cal P}_m=-{1\over 2\pi i}\int\limits_{\cal R}{dk'_m\over k'_m-(k_m-iO)}.
\eeq
Equations (\ref{eq:solwquadrn}) summarize all the analyticity 
informations contained in the ZRC, allow one to express the unknown BVs in 
terms of given data and, via (\ref{eq:Freprboxn}), to solve the Dirichelet 
problem. 

\noindent
{\it Neumann problem}. 
Similar considerations can be made in solving the Neumann BV problem. 
In this case:
\bea\label{eq:elimjN}
\ba{c}
i\chi_j({\bf k}_j,q)\Delta^{(j)}_+\hat v^{(0)}_{0j}({\bf k}_j,q)=
-\Delta^{(j)}_+(V({\bf k},q)|_{k_j=\chi_j}),~~~j=1,..,n,   \\
V({\bf k},q):=\hat f({\bf k},q)+i\hat u_0({\bf k})+
\sum\limits_{j=i}^n\hat v^{(1)}_{0j}({\bf k}_j,q)
\ea
\eea
and 
\bea\label{eq:solvquadrn}
\ba{c}
\hat v^{(0)}_{0j}({\bf k}_j,q)=i
{\cal P}^{(j)}\left(
{1\over \chi_j({\bf k}_j,q)}\Delta^{(j)}_+(V({\bf k},q)|_{k_j=\chi_j})
\right),~~~j=1,..,n.
\ea
\eea

In this multidimensional context, for the presence of the analyticity 
projectors, the 
unknown BVs in Fourier space turn out to be nonlocal expressions of the given 
data. It is however possible to show that, due to the analyticity 
properties of the involved FTs, it is not 
really necessary to apply the above analyticity projectors to construct the 
unknown BVs and the solution $u({\bf x},t)$ in configuration space. The 
strategy to avoid unpleasent nonlocalities is outlined in the next section 
and leads to a Fourier representation of the solution already obtained 
in \cite{DMS1} using the EbR approach. Therefore this strategy is  
also the way to establish the connection between the Analyticity and the EbR 
approaches.

\section{Connections between the Analyticity and the EbR approaches}

\noindent
{\it Dirichelet problem}

\noindent
We first remark that the unknown BVs can be constructed 
directly in terms of known data from the RHS of (\ref{eq:elimjD}b):
\beq\label{eq:solwquadrnbis}
v^{(1)}_{0j}({\bf x}_j,t)=
-\int_{{\cal R}^{n}}{d{\bf k}_jdq\over (2\pi )^n}
e^{i({\bf k}_j\cdot {\bf x}_j+qt)}\Delta^{(j)}_-(W({\bf k},q)|_{k_j=
\chi_j}),~~~t>0,~x_k\ge 0,~k\ne j.
\eeq
Indeed, from the analyticity properties of $\hat v^{(1)}_{0j}$ we know  
that its  inverse FTs (\ref{eq:invFT}) is zero 
outside the domain of definition in configuration space 
(i.e., for $x_k<0,~k\ne j$); this implies the formula 
\beq
\int_{{\cal R}^{n}}d{\bf k}_jdqe^{i({\bf k}_j\cdot {\bf x}_j+qt)}
[\hat v^{(1)}_{0j}({\bf k}_j,q)-\Delta^{(j)}_-\hat v^{(1)}_{0j}({\bf k}_j,q)]
=0,~t>0,~x_j>0,~j=1,..,n
\eeq
and, through (\ref{eq:elimjD}a), equation (\ref{eq:solwquadrnbis}).

Also the solution $u({\bf x},t)$ can be reconstructed without going through 
the nonlocalities associated with the analyticity projectors. Indeed it is 
possible to show that the following relation holds true:
\beq\label{eq:solwquadrnter}
\sum\limits_{j=1}^n\hat v^{(1)}_{0j}({\bf k}_j,q)~\tilde =
~(\Delta_--1)W({\bf k},q),
\eeq
where the equivalence $\hat A({\bf k},q)~\tilde =~\hat B({\bf k},q)$ means that
the FTs $\hat A({\bf k},q)$ and $\hat B({\bf k},q)$ are equal under the 
following Fourier integral projector:
\beq\label{eq:equiv}
\int\limits_{{\cal R}^{n+1}}d{\bf k}dq
{e^{i({\bf k}\cdot {\bf x}+qt)}\over q+k^2-i0} 
[\hat A({\bf k},q)-\hat B({\bf k},q)]=0,~~~({\bf x},t)\in{\cal D}.
\eeq
The equivalence (\ref{eq:solwquadrnter}) and equation (\ref{eq:elimjD}) imply
\beq
\hat{\cal N}({\bf k},q)~\tilde =~\Delta_-W({\bf k},q)
\eeq
and the following spectral representation of the solution in terms of 
known data:
\bea\label{eq:SFrepr}
\ba{c}
u({\bf x},t)=
\int_{{\cal R}^{n+1}}{dqd{\bf k} \over (2\pi )^{n+1}}
e^{i({\bf k}\cdot {\bf x}+qt)}
{\Delta_-\hat f({\bf k},q)\over q+k^2-iO}
+\int_{{\cal R}^n}{d{\bf k} \over (2\pi )^n}
e^{i({\bf k}\cdot {\bf x}-k^2t)}\Delta_-\hat u_0({\bf k})+  \\
\sum\limits_{j=1}^n\int\limits_{{\cal R}^{n-1}}{d{\bf k}_j\over 
(2\pi )^{n-1}}
(\int_{{\cal K}^{(i)}_1}{dk_j\over \pi}e^{i({\bf k}\cdot {\bf x}-k^2t)} 
k_j\Delta^{(j)}_-\hat v^{(0)}_{0j}({\bf k}_j,-k^2),~~({\bf x},t)\in{\cal D},
\ea
\eea
already obtained in \cite{DMS1} using the EbR approach.

\noindent
The proof of (\ref{eq:solwquadrnter}) is based on the important fact 
that all the admissible 
Fourier transforms $\hat A({\bf k},q)$ under consideration satisfy the 
equivalence 
\beq\label{eq:equivFT}
\hat\sigma_j\hat A({\bf k},q)~\tilde =~\hat A({\bf k},q)|_{k_j=\chi_j},
~~~~j=1,..,n
\eeq 
and goes as follows. For $n=2$, the $2$ ZRCs 
(\ref{eq:ZRCquadrn}) and their consequences (\ref{eq:elimjD}) yield the 
$4$ equivalence relations:
\bea
\ba{c}
\hat v^{(1)}_{01}(k_2,q)+\hat\sigma_1\hat v^{(1)}_{02}(k_1,q)\tilde =
-\hat\sigma_1W({\bf k},q),~~
\hat\sigma_2\hat v^{(1)}_{01}(k_2,q)+\hat v^{(1)}_{02}(k_1,q)\tilde =
-\hat\sigma_2W({\bf k},q),  \\
(1-\hat\sigma_2)\hat v^{(1)}_{01}(k_2,q)\tilde =
-\hat\sigma_1(1-\hat\sigma_2)W({\bf k},q),~~
(1-\hat\sigma_1)\hat v^{(1)}_{02}(k_1,q)\tilde =
-\hat\sigma_2(1-\hat\sigma_1)W({\bf k},q)
\ea
\eea
and their sum is exactly equation (\ref{eq:solwquadrnter}). To generalize this 
result to the case of an 
arbitrary $n$, consider the $n$ ZRCs (\ref{eq:ZRCquadrn}) and 
all their 
consequences, obtained applying systematically parity operators characterized 
by different indeces:
\bea
\ba{c}
\hat v^{(1)}_{0j}({\bf k}_j,q)+
\hat\sigma_j\sum\limits_{l\ne j}\hat v^{(1)}_{0l}({\bf k}_l,q)
~\tilde =~-\hat\sigma_jW({\bf k},q),~~~~~~j=1,..,n,  \\
(1-\hat\sigma_i)\hat v^{(1)}_{0j}({\bf k}_j,q)+
\hat\sigma_j(1-\hat\sigma_i)\sum\limits_{l\ne j}\hat v^{(1)}_{0l}({\bf k}_l,q)
~\tilde = \\
-\hat\sigma_j(1-\hat\sigma_i)W({\bf k},q),~~~i\ne j,~~j=1,..,n,  \\
...............   \\
\Delta^{(j)}_-\hat v^{(1)}_{0j}({\bf k}_j,q)~\tilde =~
-\hat\sigma_j\Delta^{(j)}_-W({\bf k},q),~~~j=1,..,n.
\ea
\eea
The sum of all these equations with weights $1/{n-1\choose m}$ ($m$ is 
the number of parity operators appearing in the equation) yields the result 
(\ref{eq:solwquadrnter}).

\noindent
{\it Neumann problem}

\noindent  
Similar considerations can be made in the case of the Neumann IBV problem. 
Now the unknown BVs are recovered via:
\beq
v^{(0)}_{0j}({\bf x}_j,t)=-\int_{{\cal R}^{n}}{d{\bf k}_jdq\over (2\pi )^n}
e^{i({\bf k}_j\cdot {\bf x}_j+qt)}(\Delta^{(j)}_+V({\bf k},q)|_{k_j=
\chi_j}),~t>0,~x_k\ge 0,~k\ne j
\eeq
and the spectral representation of the solution reads:
\bea\label{eq:CFrepr}
\ba{c}
u({\bf x},t)=-\int_{{\cal R}^{n+1}}{dqd{\bf k} \over (2\pi)^{n+1}i}
e^{i({\bf k}\cdot {\bf x}+qt)}
{\Delta_+\hat f({\bf k},q)\over q+k^2-iO}
+\int_{{\cal R}^n}{d{\bf k} \over (2\pi )^n}
e^{i({\bf k}\cdot {\bf x}-k^2t)}\Delta_+\hat u_0({\bf k})-  \\
\sum\limits_{j=1}^n\int\limits_{{\cal R}^{n-1}}{d{\bf k}_j\over 
(2\pi )^{n-1}}
(\int_{{\cal K}^{(i)}_1}{dk_j\over \pi i}e^{i({\bf k}\cdot {\bf x}-k^2t)} 
\Delta^{(j)}_+\hat v^{(1)}_{0j}({\bf k}_j,-k^2),~~({\bf x},t)\in{\cal D};
\ea
\eea
a formula already derived in \cite{DMS1} using the EbR approach.

\vskip 10pt 
\noindent
{\bf Acknowledgments}

\vskip 5pt
\noindent
The present work has been carried out during several visits and meetings. 
We gratefully acknowledge the financial contributions provided by the 
RFBR Grant 01-01-00929, the INTAS Grant 99-1782 and by the 
following Institutions: the University of Rome ``La Sapienza'' (Italy), 
the Istituto Nazionale di Fisica Nucleare (Sezione di Roma), the Landau 
Institute for 
Theoretical Physics, Moscow (Russia) and the Isaac Newton Institute,  
Cambridge (UK),  
within the programme ``Integrable Systems''.

\end{document}